\RequirePackage{fix-cm}
\documentclass[smallextended]{svjour3}       % onecolumn (second format)
\smartqed  % flush right qed marks, e.g. at end of proof
\usepackage[colorlinks,
linkcolor = red,
urlcolor  = blue,
citecolor = blue,
anchorcolor = blue]{hyperref}
\usepackage{color}
\usepackage{color,soul}
\usepackage{graphicx}
\usepackage{slashed}
\begin{document}

\title{Dissociation of Pseudotensor Mesons through Critical Temperature}

\subtitle{}


\author{A. T\"{u}rkan        \and
         J.Y.~S\"ung\"u \and E. G\"{u}ng\"{o}r \and  H. Reiso\u{g}lu \and E. Veli Veliev %etc.
}


\institute{A. T\"{u}rkan \at
              \"{O}zye\u{g}in University, Department of Natural and
              Mathematical Sciences, \c{C}ekmek\"{o}y,Istanbul,Turkey \\
              Tel.:+90(216) 564 90 00\\
              Fax:+90(216) 564 99 99\\
              \email{arzu.turkan@ozyegin.edu.tr},           %  \\
           \and         
           J.Y.~S\"ung\"u, E. G\"{u}ng\"{o}r, H. Reiso\u{g}lu, E. Veli Veliev  \at
             Department of Physics, Kocaeli University, 41001
             Izmit, Turkey}

\date{Received: date / Accepted: date}

\maketitle

\begin{abstract}
We have computed the masses and decay constants of isotriplet
 $\pi_2(1670)$, the isoscalar $\eta_2(1645)$ and 
$\eta_2(1870)$ states as the
ground-state nonet of $1^1D_2$ case within the Thermal QCD sum
rules framework. This method is applied to the spectral changes of
pseudotensor mesons at high temperatures including quark, gluon, and mixed condensates up to the five dimensions. We found that their masses and decay constants are insensitive to temperature in low-temperature region. At the near critical temperature, we observed an exponential
decrease of the masses of $\pi_2$, $\eta_2(1645)$, and $\eta_2(1870)$, but while the decay constant of $\pi_2$ is fallen, that of $\eta_2(1645)$ and $\eta_2(1870)$ is increased as a function of
temperature. The modification of condensates at $T\neq 0$ is crucial to the change of considered
mesons properties in hot medium.  We discuss and interpret the implications of the
physical meaning of our results. 
\end{abstract}

\section{Introduction}

Conventional quark-antiquark meson states are classified as scalar
$(1^3P_0,~J^{PC} = 0^{++})$, pseudoscalar $(1^1S_0,~J^{PC} =
0^{-+})$, vector $(1^3S_1,~J^{PC} = 1^{--})$, axial-vector
$(1^3P_1,~J^{PC}= 1^{++}, 1^1P_1,~J^{PC} =
1^{+-})$, pseudotensor $(1^ 1D_2,~J^{PC} = 2^{-+},1^3D_2,~J^{PC} =2^{--})$ and
tensor $(1^3P_2,~J^{PC} =2^{++})$. Among them, the tensor
meson nonet is not well established \cite{Wang:2014sea} and there aren't many studies in the literature, but it has been one of the issues that attracted attention in recent years.

$\pi_2(1670)$ was first detected at Aachen-Berlin-CERN
collaboration in the reaction $\pi^-p \rightarrow p
\pi^+\pi^-\pi^-$. Its mass and width are found to be $(1660 \pm
16)$ MeV and $(270 \pm 60)$ MeV~\cite{Ascoli:1973}, respectively.
Another family member of the pseudotensor mesons, the
$\eta_2(1870)$ has been observed in $\gamma\gamma$ reactions
~\cite{Karch:1991sm}, $p\bar{p}$
annihilation~\cite{Anisovich:2000mv}, and radiative $J/\psi$ decays
in BES Collaboration~\cite{Bai:1999tg}. This state has been
detected by the Crystal Ball collaboration~\cite{Karch:1991sm} in
the final state $\eta\pi^0\pi^0$ of a $\gamma\gamma$ reaction as a
resonant structure finding mass $(1881\pm 32\pm40)$ MeV and width
$(221\pm92\pm44)$ MeV, and in the reaction
$p\bar{p}\longrightarrow\eta\pi^0\pi^0\pi^0$ by the Crystal Barrel
collaboration, with mass $(1875\pm20\pm25)$ MeV, and width 
$(200\pm25\pm45)$ MeV and for the $\eta_2(1645)$ state,
$(1645\pm14\pm15)$ MeV, $(180^{+40}_{-21}\pm25)$ MeV,
respectively~\cite{Adomeit:1996nr}. 

The study of the Regge Trajectory model indicates that both
$\eta_2{(1645)}$ and $\eta_2{(1870)}$ are the eigenvectors of a mass-squared matrix, the physical isoscalar states can be related to the bare states as:\\
\begin{equation}\label{MixingMatrix}
	\left(\begin{array}{c}
		|\eta_2{(1645)}\rangle\\
		|\eta_2{(1870)}\rangle
	\end{array}\right)=\left(\begin{array}{cc}\cos\theta&\sin\theta \\
		-\sin\theta&\cos\theta\end{array}\right)\left(\begin{array}{c}\Psi_1
		\\\Psi_8\end{array}\right),
\end{equation}
\begin{eqnarray}\label{Psi1and8}
	~~~~~~~~\Psi_1&=&\frac{1}{\sqrt{3}}(u\bar{u}+d\bar{d}+s\bar{s}),\nonumber\\ 	~~~~~~~~\Psi_8&=&\frac{1}{\sqrt{6}}(u\bar{u}+d\bar{d}-2s\bar{s}),
\end{eqnarray}
where the mixing angle $\theta$ is an important input
parameter determined by the experimental data. From now on, we will use $\eta_2^{(1)} $ and $\eta_2^{(2)} $ symbols for the $ \eta_2{(1645)} $ and $ \eta_2{(1870)} $ mesons, respectively in text for brevity. Although we have no clear information on the mixing angle of these particles~\cite{Zyla:2020zbs}, this angle is estimated to be small. So, we assume that there is no mixing between them.

In this study, we use the interpolating current associated with quantum numbers
$J^{PC}=2^{-+}$ to evaluate the two-point correlation function at
non-zero temperatures. We employ the Thermal QCD sum rules (TQCDSR) technique for analyses and
extract the masses and decay constants sum rules of the considered
pseudotensor states depending on temperature. This method has been applied successfully for the tensor mesons in the literature~\cite{Turkan:2019anj,Sungu:2020azn}. A comparison of our
results with predictions at vacuum obtained from other approaches
is also made. These kinds of works are helpful for both fulfilling
the hadron spectrum and allowing us to achieve some hints about
quark-gluon plasma (QGP) which is predicted to cover the universe
at the early stage of the universe. Quarks and gluons condense
to create a gas of nucleons and light mesons, the latter decay subsequently. Jet
quenching for light mesons at RHIC
is well explained by radiative energy loss calculations ~\cite{Vitev:2005he,Lajoie:2006ha}. As a result, probing the variations of hadronic parameters depending on temperature can provide us valuable information on the new phase of matter, i.e. QGP.

The outline of the article is arranged as follows. In
Section~\ref{sec:SR}, we shortly describe the TQCDSR method used in the computation. The masses and decay
constants of the light-unflavored $\pi_2$, $\eta_2^{(1)}$,and $\eta_2^{(2)}$ mesons are
extracted from the TQCDSR. In the following, numerical analysis is introduced in Section~\ref{sec:NumAnal}. Next, the numerical results
will be discussed in Section~\ref{sec:Result} and compared the
results with other models and lastly, the explicit forms of thermal spectral functions are given in Appendix ~\ref{sec:App}.
%%%%%%%%%%%%%%%%%%%%%%%%%%%%%%%%%%%%%%%%%%%%%%%%%%%%%%%%%%%%%%%%%%%%%%%%%%%%%
\section{The Methodology\label{sec:SR}}
%%%%%%%%%%%%%%%%%%%%%%%%%%%%%%%%%%%%%%%%%%%%%%%%%%%%%%%%%%%%%%%%%%%%%%%%%%%%%

The TQCDSR approach is a beautiful tool to describe
the variations of the hadronic parameters (mass, coupling constant, etc.) according to thermal expectation values of quark and gluon condensates
(or light-cone distribution amplitudes) at low energy~\cite{Bochkarev:1985ex}. TQCDSR use dispersion representations to compute QCD
Green functions, the time-ordered $(\mathcal{T})$ products of
local hadron interpolating currents built of quark and gluon
fields, in two different schemes: These are the QCD part (or named as theoretical or OPE side) which is defined in terms of quark and gluon degrees of freedom, and the hadronic part (called the physical or phenomenological side) related to hadronic degrees of freedom. The QCD side is computed by converting the time-ordered product into a sum of local operators, i.e thermal quark and gluon condensates via Wilson expansion (OPE) separating systematically short and long-distance effects. Thermal quark and gluon condensates are nonperturbative in nature and seem to power corrections to
the leading logarithmic, namely perturbative behaviour. They have greater importance than higher-order $\alpha_s$ corrections and leading logarithmic contributions ~\cite{Shifman:1978bx,Shifman:1978by}.

Thermal condensates parametrize the nonperturbative properties of the QCD in hot medium. However, at finite temperatures, additional operators arise in the rest frame of the heat bath. They can be expressed as Lorentz scalars in a general frame by employing the four-velocity vector $u_{\mu}$. They come out in 4-dimension, which are  $u^{\mu}\Theta^f_{\mu\nu} u^{\nu} $ and $u^{\mu} \Theta^g_{\mu\nu} u^{\nu} $, where $\Theta^{f,g}_{\mu\nu}$ are the fermionic and gluonic parts of energy-momentum tensors of QCD, respectively~\cite{Mallik:1997kj}. We will mention this later in detail in the numerical illustration section.

We first begin to consider the usual current correlator to the computations~\cite{Bochkarev:1985ex}-\cite{Shifman:1978by}:
\begin{equation}\label{eq:CorrFunc}
\Pi_{\mu \nu,\alpha\beta }(q,T)=i\int d^{4}x~e^{iq\cdot (x-y)}\langle
\varrho \mathcal{T} \big(J_{\mu\nu}^{PT}(x)J_{\alpha\beta}^{\dagger PT}(y)\big) \rangle,
\end{equation}
here $ PT $ symbolizes the pseudotensor states $\pi_2$, $\eta_2^{(1)}$ and $\eta_2^{(2)}$, $ T$ is temperature, $\mathcal{T}$ represents the time-ordered product and 
\begin{equation}\label{eq:density}
\varrho=\frac{e^{-H/T}}{Tr(e^{-H/T})},
\end{equation}
is the thermal density matrix of QCD. 

In this paper, we restrict ourselves to the currents with
quantum numbers $J^{PC}= 2^{-+}$ for the considered pseudotensor
states $ \pi_2,\eta_2^{(1)} $ and $\eta_2^{(2)} $, respectively ~\cite{Reinders:1984sr}:
\begin{eqnarray}\label{eq:Current}
J^{\pi_2}_{\mu\nu}&=&\frac{i}{2\sqrt{2}}\bigg\{\bar{u}_a\left({\gamma_{\mu}}{\gamma_{5}}
\partial_{\nu}+{\gamma_{\nu}}{\gamma_{5}}\partial_{\mu}+{\frac{2}{3}}{\eta_{\mu\nu}}{\gamma_{5}}\slashed{\partial}\right){u_{a}}\nonumber \\&-&\bar{d}_a\left({\gamma_{\mu}}{\gamma_{5}}
\partial_{\nu}+{\gamma_{\nu}}{\gamma_{5}}\partial_{\mu}+{\frac{2}{3}}{\eta_{\mu\nu}}{\gamma_{5}}\slashed{\partial}\right){d_{a}}\bigg\},\nonumber \\
J^{\eta_2^{(1)}}_{\mu\nu}&=&\frac{i}{2\sqrt{3}}\bigg\{\bar{u}_a\left({\gamma_{\mu}}{\gamma_{5}}\partial_{\nu}+{\gamma_{\nu}}{\gamma_{5}}\partial_{\mu}+{\frac{2}{3}}{\eta_{\mu\nu}}{\gamma_{5}}\slashed{\partial}\right){u_{a}}\nonumber \\
&+&\bar{d}_a\left({\gamma_{\mu}}{\gamma_{5}}
\partial_{\nu}+{\gamma_{\nu}}{\gamma_{5}}\partial_{\mu}+{\frac{2}{3}}{\eta_{\mu\nu}}{\gamma_{5}}\slashed{\partial}\right){d_{a}}\nonumber \\
&+&\bar{s}_a\left({\gamma_{\mu}}{\gamma_{5}}
\partial_{\nu}+{\gamma_{\nu}}{\gamma_{5}}\partial_{\mu}+{\frac{2}{3}}{\eta_{\mu\nu}}{\gamma_{5}}\slashed{\partial}\right){s_{a}}\bigg\},\nonumber \\
J^{\eta_2^{(2)}}_{\mu\nu}&=&\frac{i}{2\sqrt{6}}\bigg\{\bar{u}_a\left({\gamma_{\mu}}{\gamma_{5}}
\partial_{\nu}+{\gamma_{\nu}}{\gamma_{5}}\partial_{\mu}+{\frac{2}{3}}{\eta_{\mu\nu}}{\gamma_{5}}\slashed{\partial}\right){u_{a}}\nonumber \\
&+&\bar{d}_a\left({\gamma_{\mu}}{\gamma_{5}}
\partial_{\nu}+{\gamma_{\nu}}{\gamma_{5}}\partial_{\mu}+{\frac{2}{3}}{\eta_{\mu\nu}}{\gamma_{5}}\slashed{\partial}\right){d_{a}}\nonumber \\
&-&2\bar{s}_a\left({\gamma_{\mu}}{\gamma_{5}}
\partial_{\nu}+{\gamma_{\nu}}{\gamma_{5}}\partial_{\mu}+{\frac{2}{3}}{\eta_{\mu\nu}}{\gamma_{5}}\slashed{\partial}\right){s_{a}}\bigg\},
\end{eqnarray}
here, $a$ is the color index, and $ \eta_{\mu\nu} $ is:
\begin{equation}\label{eq:eta}
\eta_{\mu\nu}=\frac{q_{\mu} q_{\nu}}{q^2 }-g_{\mu\nu}.  
\end{equation}

At the hadron level, i.e.~in the physical side,  a complete set of intermediate hadronic states, which has the same quantum numbers as the current operator is embedded into the correlator to get the hadronic representation. Then, subtracting the ground state contribution from other states and taking the integration by $x$ and setting $ y = 0 $, we can obtain the physical side of the correlator as follows and from now on we shortly call it the left-hand side (LHS) in formulae:
\begin{eqnarray}\label{eq:Phys1}
\Pi _{\mu \nu, \alpha\beta }^{\mathrm{LHS}}(q,T)&=&\frac{\langle \Omega|J_{\mu\nu}^{PT}|PT\rangle \langle PT|J_{\alpha\beta }^{\dagger PT}| \Omega \rangle}{m_{PT}^{2}(T)-q^{2}}+\ldots,
\end{eqnarray}
where $m_{PT}(T)$ shows the temperature-dependent masses of pseudotensor states handled. $ \Omega $ indicates the heat bath and dots symbolize the higher resonances and continuum states.

To continue the computation of the physical side's thermal sum rules,
we present the matrix elements through masses and decay
constants of pseudotensor mesons. The decay constants and matrix element of the pseudotensor current between the one meson state and the thermal medium are related to each other as follows:
\begin{eqnarray}\label{eq:Matrixel}
&&\langle \Omega|J_{\mu\nu}^{PT}|PT\rangle=f_{PT}(T) m_{PT}^3 (T)\varepsilon_{\mu\nu}.
\end{eqnarray}
In Eq.~(\ref{eq:Matrixel}), $\varepsilon _{\mu\nu }$ is the polarization tensor of the
considered mesons and obey the following relation:
\begin{equation}\label{eq:Poltensor}
\varepsilon_{\mu\nu}\varepsilon^{*}_{\alpha\beta}=\frac{1}{2}{T_{\mu\alpha}}{T_{\nu\beta}}+\frac{1}{2}{T_{\mu\beta}}{T_{\nu\alpha}}-{\frac{1}{3}}{T_{\mu\nu}}{T_{\alpha\beta}},
\end{equation}
\begin{equation}\label{eq:eta}
T_{\mu\nu}=\frac{q_{\mu} q_{\nu}}{m^2_{PT} (T)}-g_{\mu\nu}.  
\end{equation}
In our calculations, we select following structure in $\Pi _{\mu \nu,\alpha\beta}^{\mathrm{LHS}}(q,T)$:
\begin{eqnarray}\label{eq:Phys2}
\Pi_{\mu \nu,\alpha\beta}^{\mathrm{LHS}}(q,T)=\Pi ^{\mathrm{LHS}}(q^{2},T)\Bigg[\frac{1}{2} (g_{\mu \alpha} g_{\nu \beta}+ g_{\mu \beta} g_{\nu \alpha }) \Bigg]+other~structures,~
\end{eqnarray}
here $ \Pi ^{\mathrm{LHS}}(q^{2},T)= \frac{m_{PT}^{6}(T)f_{PT}^{2}(T)}{m^2_{PT}(T)-q^2}$ represents the scalar function belonging to the chosen structure. In the next step, the Borel transformation is applied to Eq.~(\ref{eq:Phys2}) and only the coefficient of the selected structure is written as:\\
\begin{eqnarray}\label{eq:Phys3}
B(q^2)\Pi^{\mathrm{LHS}}(q^2,T)&=&m_{PT}^{6}(T)f_{PT}^{2}(T)e^{-m_{PT}^{2}(T)/{M}^{2}},
\end{eqnarray}\\
where $M^{2}$ is the arbitrary Borel mass parameter in the theory.

The QCD side of the thermal sum rule should be expressed in terms of the quark propagators.  Note that it will be called as right-hand side (RHS) with a superscript on the correlator hereafter. The currents given in Eq.~(\ref{eq:Current}) are substituted in Eq.~(\ref{eq:CorrFunc}) and the quark fields are contracted using Wick theorem.  After some straightforward calculations, we obtain the below expressions for the considered states:\\
\begin{eqnarray}\label{eq:CorrFunc2}
\Pi_{\mu \nu, \alpha\beta}^{\mathrm{RHS(\pi_2)}}(q,T) &=&-\frac{3i}{8}\int d^{4}x~e^{iq\cdot (x-y)}~\bigg\{\Gamma_{u}+\Gamma_{u \rightarrow d}\bigg\},\nonumber \\
\Pi_{\mu \nu, \alpha\beta}^{\mathrm{RHS(\eta_2^{(1)})}}(q,T) &=&-\frac{i}{4}\int d^{4}x~e^{iq\cdot( x-y)}~\bigg\{\Gamma_{u}+\Gamma_{u \rightarrow d}+\Gamma_{u \rightarrow s} \bigg\}, \nonumber \\
\Pi_{\mu \nu, \alpha\beta}^{\mathrm{RHS(\eta_2^{(2)})}}(q,T) &=&-\frac{i}{8}\int d^{4}x~e^{iq\cdot (x-y)}~\bigg\{\Gamma_{u}+\Gamma_{u \rightarrow d}+4\Gamma_{u \rightarrow s} \bigg\}, 
\end{eqnarray}
\begin{eqnarray}\label{eq:Gamau}
\Gamma_{u}&=&\bigg(\mathrm{Tr}~\big[\partial_{\beta}(y)S_{u}(y-x)\gamma_{\mu }\gamma_{5}\partial_{\nu}(x)S_{u}(x-y)\gamma_{5}\gamma_{\alpha}\big] \nonumber \\
&+&(\beta\rightarrow\alpha)+(\mu\rightarrow\nu)+(\beta\rightarrow\alpha,\mu\rightarrow\nu)\bigg)\nonumber\\
&+&\frac{2}{3}\eta_{\alpha\beta}\bigg(\mathrm{Tr}~\big[\partial_{\phi}(y)S_{u}(y-x)\gamma_{\mu }\gamma_{5}\partial_{\nu}(x)S_{u}(x-y)\gamma_{\phi}\gamma_5\big] +(\mu\rightarrow\nu)\bigg)\nonumber\\
&+&\frac{2}{3}\eta_{\mu\nu}\bigg(\mathrm{Tr}~\big[\partial_{\beta}(y)S_{u}(y-x)\gamma_{5}\gamma_{\theta}\partial_{\theta}(x)S_{u}(x-y)\gamma_5\gamma_{\alpha}\big] +(\beta\rightarrow \alpha)\bigg)\nonumber\\
&+&\frac{4}{9}\eta_{\mu\nu}\eta_{\alpha\beta}\bigg(\mathrm{Tr}~\big[\partial_{\phi}(y)S_{u}(y-x)\gamma_{5}\gamma_{\theta}\partial_{\theta}(x)S_{u}(x-y)\gamma_{\phi}\gamma_5\big] \bigg).
\end{eqnarray}
Then we will set $ y = 0 $ after applying derivative with respect to $ y $. In Eq.~(\ref{eq:CorrFunc2}) $S_q^{ij}$ is the thermal light quark propagator~\cite{Mallik:1997pq}:
\begin{eqnarray}\label{eq:lightquarkpropagator}
	S_{q}^{ij}(x-y) &=&i\frac{\slashed
		x-\slashed y}{2\pi^{2}(x-y)^{4}}\delta_{ij}-\frac{
		m_{q}}{4\pi^{2}(x-y)^{2}}\delta_{ij} -\frac{\langle \bar{q}q\rangle_T }{12}\delta_{ij}-\frac{(x-y)^{2}}{192}%
	m_{0}^{2}\langle \bar{q}q\rangle_T \nonumber \\
	&\times& \Big[1-i\frac{m_{q}}{6}(\slashed x-\slashed y) \Big]%
	\delta _{ij} +\frac{i}{3}\Big[(\slashed x-\slashed y) \Big(\frac{m_{q}}{16}\langle
	\bar{q}q\rangle_T -\frac{1}{12}\langle u^{\mu} \Theta _{\mu \nu
	}^{f} u^{\nu}\rangle \Big) \nonumber \\&+& \frac{1}{3}\Big(u\cdot (x-y)\Big)\slashed u
	\langle u^{\mu}\Theta _{\mu \nu
	}^{f} u^{\nu}\rangle %
	\Big]\delta _{ij}  -\frac{ig_{s}G _{ij}^{\alpha\beta}}{32\pi ^{2}(x-y)^{2}}\nonumber \\
	&\times&
	\Big((\slashed x-\slashed y)\sigma _{\alpha\beta }+\sigma _{\alpha\beta }(\slashed x-\slashed y)\Big)-i\delta_{ij}\frac{(x-y)^2 (\slashed x-\slashed y) \langle \bar{q}q
		\rangle^2_T}{7776} g_s^2,
\end{eqnarray}\\
here we use the notation $ G_{ij}^{\alpha \beta }=G_{A}^{\alpha\beta}t_{ij}^{A} $, $A=1,\,2\,\ldots 8$ are gluon color indices, $t^{A}=\lambda^{A}/2$ with Gell-Mann matrices
$\lambda^{A}$, and the gluon field strength tensor $G_{\alpha \beta}^{A}\equiv
G_{\alpha \beta }^{A}(0)$ is fixed at $x=0$.

The TQCDSR of the related pseudotensor states can be extracted equalizing the same structures in both $\Pi_{\mu \nu, \alpha \beta}^{\mathrm{LHS}}(q,T)$ and $\Pi _{\mu \nu, \alpha \beta  }^{\mathrm{RHS}}(q,T)$. The expression of the light quark propagator is inserted in Eq.~(\ref{eq:CorrFunc2}), the derivatives are taken and $ y $ is set to zero, after calculating the integrals, we obtain the explicit tensor structure of $\Pi_{\mu \nu, \alpha \beta  }^{\mathrm{RHS}}(q,T)$. To continue our calculations, the above structures in Eq.~(\ref{eq:Phys2}) are chosen from the $\Pi_{\mu \nu, \alpha \beta  }^{\mathrm{RHS}}(q,T)$ and then we get the related scalar function $  \Pi^{\mathrm{RHS}}(q^{2},T) $ which can be separated as a sum of perturbative and non-perturbative terms:
\begin{equation}\label{eq:RHS}
\Pi^{\mathrm{RHS}}(q^{2},T)=\Pi^{Pert.}(q^{2},T)+\Pi^{Non-Pert.}(q^{2},T).
\end{equation}
The expression of spectral density $ \rho $  for the perturbative part is obtained, which is given by the dispersion relation as follows:\\
\begin{equation}\label{eq:PiQCD}
\Pi^{\mathrm{RHS}}(q^{2},T)=\int_{s_{min}}^{\infty}ds~\frac{\rho^{Pert.}(s,T)}{s-q^{2}}+\mathrm{subtracted~terms},
\end{equation}
\begin{equation}\label{eq:rhoRHS}
\rho^{Pert.}(s,T)=\frac{1}{\pi}Im [\Pi^{RHS}(s,T)].
\end{equation}\\
In Eq.~(\ref{eq:PiQCD}),  $s_{min}$ is determined with the masses of quarks of the related meson. The TQCDSR approach connects the perturbative and nonperturbative
territory with the assumption of quark-hadron duality. To eliminate the contribution coming from the higher
resonances and continuum states, the Borel transformation is performed. After calculating the non-perturbative terms with the direct Borel method, the effective mass and decay constant of the related states are obtained as:\\
\begin{equation}\label{eq:MassSR}
m_{PT}(T)=\sqrt{\frac{\int_{s_{min}}^{s_{0}(T)}ds~\rho^{Pert.}(s)~s~e^{-s/M^{2}}+\frac{d}{d(-1/M^2)}B\Pi^{Non-Pert.}(q,T)}{\int_{s_{min}}^{s_{0}(T)}ds~\rho^{Pert.}(s)~e^{-s/M^{2}}+B\Pi^{Non-Pert.}(q,T)}},
\end{equation}
\begin{eqnarray}\label{eq:DecConSR}
f_{PT}(T)=\sqrt{\frac{\int_{s_{min}}^{s_{0}(T)}ds~\rho^{Pert.}(s)~e^{-s/M^2}+B\Pi^{Non-Pert.}(q,T)}{m_{PT}^6(T)~e^{-m^2_{PT}(T)/M^2}}}.
\end{eqnarray}
The explicit expressions of perturbative spectral density term and Borel transformed non-perturbative contributions are presented in the Appendix \ref{sec:App}. In Eq.~(\ref{eq:MassSR}) and Eq.~(\ref{eq:DecConSR}) there are two free parameters in the models, which should be chosen to obtain the stable thermal sum
rules: the Borel parameter $M^{2}$  which establishes convergence of the OPE and the cut-off parameter $s_{0}$  isolate the contribution of ground state from the higher resonances and continuum.	
%%%%%%%%%%%%%%%%%%%%%%%%%%%%%%%%%%%%%%%%%%%%%%%%%%%%%%%%%%%%%%%%%%%%%%%%%%%%
\section{Numerical Illustration}\label{sec:NumAnal}
%%%%%%%%%%%%%%%%%%%%%%%%%%%%%%%%%%%%%%%%%%%%%%%%%%%%%%%%%%%%%%%%%%%%%%%%%%%%
	
	In this part, finite temperature effects on the
	pseudotensor meson properties will be discussed and analyzed numerically in TQCDSR approach. By handling analytic results in Eq.~(\ref{eq:MassSR}) and Eq.~(\ref{eq:DecConSR}), we try to define the temperature-dependent variations of the hadronic parameters belonging to the considered states. We used the following input parameters for the quark masses, $m_{u}=2.16^{+0.49}_{-0.26}~\mathrm{MeV}$, $m_{d}=4.67^{+0.48}_{-0.17}~\mathrm{MeV}$, $m_{s}=93^{+11}_{-5}~\mathrm{MeV}$ ~\cite{Zyla:2020zbs,Narison:2010cg} and we assume the critical temperature value $T_c=155 
	~\mathrm{MeV}$ for the phase transition to
	 QGP as predicted in Refs.~\cite{Aoki:2006br}-\cite{Bazavov:2017dus}. For the $  u $ and $ d $ quarks vacuum condensate, we employ $\langle 0| \bar{q}q |0\rangle=(-275(5))^3~\mathrm{MeV}^3$ and for the $ s $ quark vacuum condensate
	 $\langle  0| \bar{s}s  |0\rangle=(-296(11))^3~\mathrm{MeV}^3$ \cite{Gubler:2018ctz}.
	
	Additionally, during our calculations, we will take into account the temperature dependent quark condensate. The below fit function for the thermal version of quark condensate is obtained from Ref.~\cite{Gubler:2018ctz,Dominguez:2016roi} by fitting
	lattice data and is defined via light quark
	vacuum condensate $\langle 0|\bar{q}q|0\rangle$:
	\begin{eqnarray}\label{eq:qbarqT}
		\langle\bar{q}q\rangle_{T}=
		(\kappa_1 e^{c_1 T}+\kappa_2)\langle 0| \bar{q}q |0\rangle,
	\end{eqnarray}
	here $q$ symbolizes the $u$ and $d$ quarks. Whereas for the $ s $ quark condensate:
	\begin{eqnarray}\label{eq:sbarsT}
		\langle\bar{s}s\rangle_{T}=
		(\kappa_3 e^{c_2 T}+\kappa_4)\langle 0| \bar{s}s |0\rangle,
	\end{eqnarray}
	with $ c_1=\mathrm{0.040~MeV^{-1}}$, $c_2=\mathrm{0.516~MeV^{-1}}$,
	$\kappa_1$$=-6.534\times10^{-4}$, $\kappa_2=1.015$,
	$\kappa_3=-2.169\times10^{-5}$, $\kappa_4$$=1.002$ \cite{Azizi:2019cmj}.
	To carry on the calculation, the temperature-dependent continuum threshold for the handled pseudotensor states also needs to be specified. The continuum threshold expression is
	generated by~\cite{Dominguez:2016roi,Borsanyi:2010bp,Bhattacharya:2014ara}:
	\begin{eqnarray}\label{eq:s0qqbarratio}
		\frac{s_0(T)}{s_0(0)}= \bigg[\frac{\langle \bar{q}q\rangle_T}{\langle 0| \bar{q}q |0\rangle }\bigg]^{2/3}.
	\end{eqnarray}
	
	The hadronic quantities rely on the Borel mass $M^2$ and cut-off parameter $s_0$ in TQCDSR. However, the selection of these free parameters has to satisfy standard restrictions. Also, we can not say the $s_0$ is totally arbitrary because it is connected with the energy of the first excited state of the considered particle. $s_0$ range is determined from the condition that assures the sum rules to have the best stability in the chosen $M^{2}$ range. Maximum and minimum bounds of $M^{2}$ are settled analysing the pole contribution. To make a reliable thermal sum rule analysis, we have to follow two criteria to determine suitable working ranges for the two auxiliary parameters $M^2$ and $s_0$:
	
	$ \bullet $ Pole dominance of the OPE; to determine the upper boundary of $M^{2}$, the pole contribution has to form more than half of the full part. This condition makes us certain that the contribution of continuum to the correlator should not be large from the pole one, i.e. pole contribution (PC) can be written as follows:
	\begin{equation}\label{eq:PC}
		\mathrm{PC}(s_0, M^2)=\frac{\Pi (M_{\mathrm{max}}^{2},\ s_{0})}{\Pi (M_{\mathrm{max}}^{2},\infty)}>\frac{1}{2}.
	\end{equation}
	
	$ \bullet $ Convergence of the OPE; to get the lowest boundary of $M^2$, we can use the usual convergence criterion as in the traditional QCD sum rules so that the contribution of highest-order operators of any dimension is less than the $\sim20\%$ of the sum of all OPE terms. We found contribution of five-dimensional mixed  condensate $ \Pi ^{(\mathrm{Dim5})} $ is:
	\begin{equation}\label{eq:Dim6Contr}
		\frac{\Pi ^{(\mathrm{Dim5})}(M_{\mathrm{min}}^{2},\ s_{0})}{\Pi^{Total} (M_{\mathrm{min}}^{2},\ s_{0})}\cong 0.002.
	\end{equation}
	
	In addition to this criteria, derived quantities should have the slightest dependence on $M^{2}$ while the chosen of $s_{0}$ interval is another constraint that has to be imposed. As a result of the analysis made by the above-mentioned conditions, our analysis yields to the below-working regions as in Table~\ref{tab:M2s0} for the $M^{2}$ and $s_{0}$ belonging to the considered pseudotensor states.
		\setlength{\tabcolsep}{0.8em} % for the horizontal padding
	{\renewcommand{\arraystretch}{1.5}% for the vertical padding
	\begin{table}[htbp]
		\caption{The working ranges of Borel mass $M^2$ and cut-off parameter $s_0$.}
		\label{tab:M2s0}
		\begin{center}
			\begin{tabular}{cccc}
				\hline
				Parameter~$[\mathrm{GeV}^2 ]$& $\pi_2(1670)$ & $\eta_2(1645)$  &$\eta_2(1870)$ \\
				\hline
				$M^2$                     & 1.6-1.8      & 1.4-1.8      & 1.8 -2.0  \\
				$s_{0}$                   & 4-4.7        &  4.0-4.6     & 5.0- 5.6  \\
				\hline
			\end{tabular}
		\end{center}
	\end{table}
	
	In the next step, using the defined free parameters, we arrive at the numerical results of our calculations at $ T=0 $ and present them in Table \ref{tab:MassResults} and \ref{tab:DecConResults}. It is seen from these Tables that
	all masses is approximately consistent with the experimental data. 	Theoretical errors in for $m_{PT}$ and $f_{PT}$ arising from uncertainties of the auxiliary parameters as well as errors in input parameters stay inside the acceptable domain for theoretical errors in the model.
	\setlength{\tabcolsep}{0.8em} % for the horizontal padding
	{\renewcommand{\arraystretch}{1.7}% for the vertical padding
	\begin{table}[htbp]
		\caption{The masses of the pseudotensor mesons with $J^{PC} = 2^{-+}$ at $ T=0 $.} \label{tab:MassResults}
		\begin{center}
			\begin{tabular}{cccc}
				\hline
				Mass~[$ \mathrm{GeV} $]& $m_{\pi_2(1670)}$ & $m_{\eta_2(1645)}$&  $m_{\eta_2(1870)}$  \\
				\hline
				Experiment~\cite{Zyla:2020zbs} & $1.671^{+2.9}_{-1.2}$ & $1.617\pm0.05$ & $1.773\pm0.08$  \\
				This~Work  &  $1.659^{+0.063}_{-0.062}$   &   $1.629^{+0.07}_{-0.007}$        &  $1.809^{+0.053}_{-0.052}$  \\
				\hline
			\end{tabular}
		\end{center}
	\end{table}
	\setlength{\tabcolsep}{0.8em} % for the horizontal padding
	{\renewcommand{\arraystretch}{1.7}% for the vertical padding
	\begin{table}[htbp]
		\caption{The decay constants of the pseudotensor mesons with $J^{PC} = 2^{-+}$ at $ T=0 $.} \label{tab:DecConResults}
		\begin{center}
			\begin{tabular}{cccc}
				\hline
				Decay constant & $f_{\pi_2(1670)}$ & $f_{\eta_2(1645)}$ & $f_{\eta_2(1870)}$  \\
				\hline
				This~Work & $6.50^{ +0.01}_{-0.02}\times 10^{-2}$   & $6.54^{+0.08}_{-0.07}\times 10^{-2} $ & $6.49^{+0.01}_{-0.01}\times 10^{-2}$   \\
				Other Theories                   & -                  & -                  &-         \\
				\hline
			\end{tabular}
		\end{center}
	\end{table} 

 Also, we tested the masses and decay constants of pseudotensor mesons are whether stable against variations at $M^2$ and $s_{0}$ and give the plot only belonging to $\pi_2(1670)$ as an example in Figure~\ref{fig2}. 
\begin{figure}[h!]
	\begin{center}
		\includegraphics[totalheight=6cm,width=7.3cm]{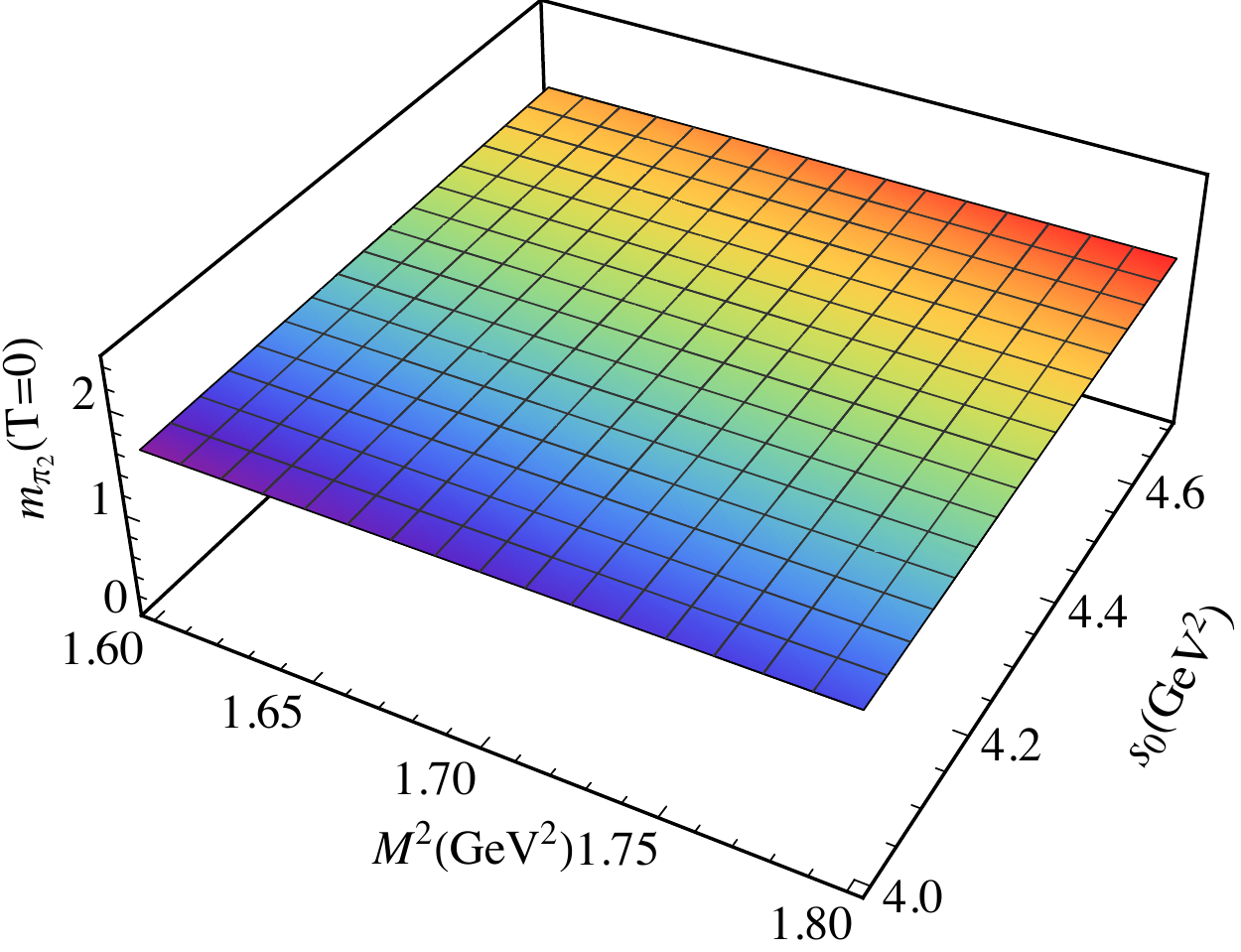}
	\end{center}
	\caption{The 3D plot of the mass of $\pi_2$ state against Borel parameter $M^2$ and continuum threshold parameter $s_0$ at $ T=0 $.} \label{fig2}
\end{figure} 

Drawing the below Figure~\ref{figmassandf}, we clearly see the deviation of hadronic parameters of pseudotensor mesons $\pi_2(1670)$, $\eta_2(1645)$ and $\eta_2(1870)$  on critical temperature at fixed $M^2$ and $s_0$. 
\begin{figure}[h!]
\begin{center}
\includegraphics[totalheight=5.2cm,width=5.8cm]{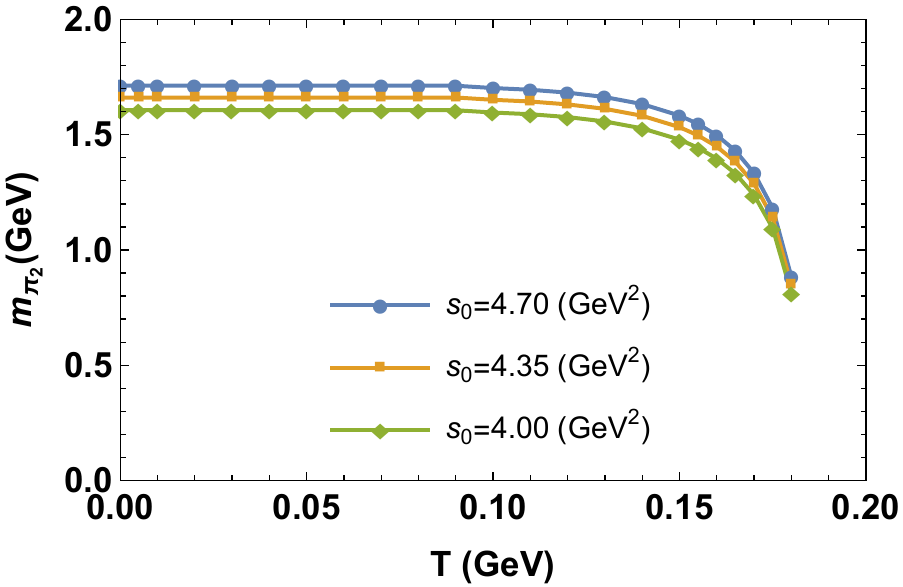}\hskip0.1cm
\includegraphics[totalheight=5.2cm,width=5.8cm]{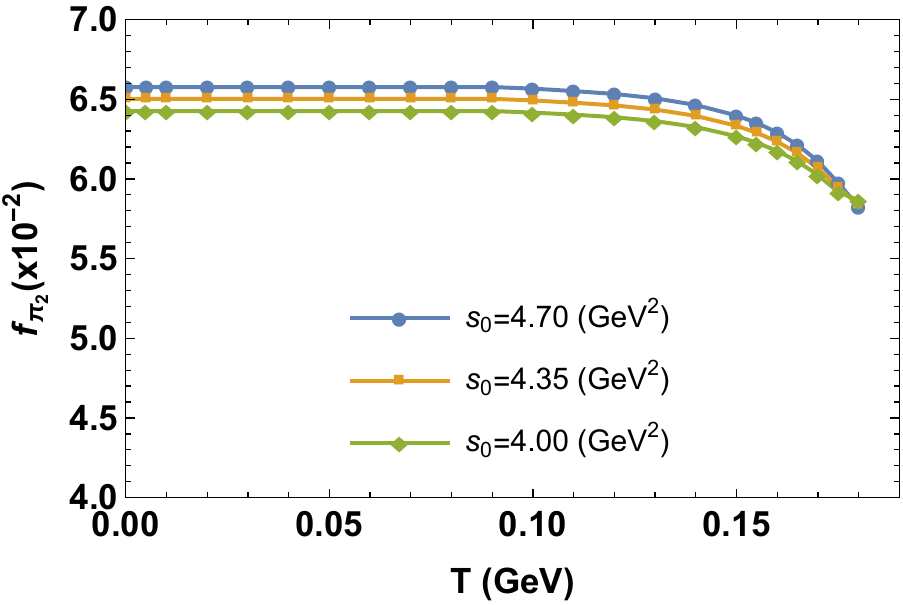}\hskip0.1cm
\includegraphics[totalheight=5.2cm,width=5.8cm]{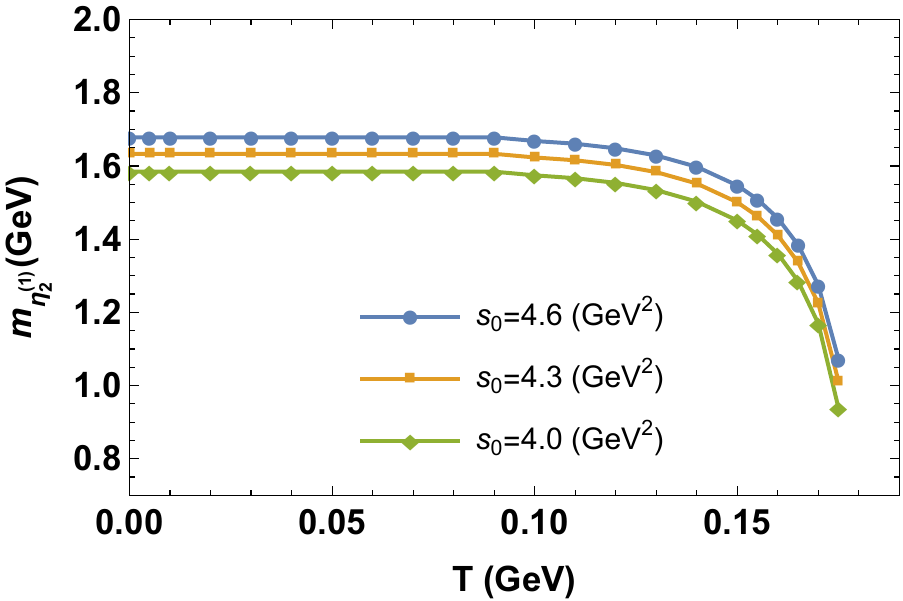}\hskip0.1cm
\includegraphics[totalheight=5.2cm,width=5.8cm]{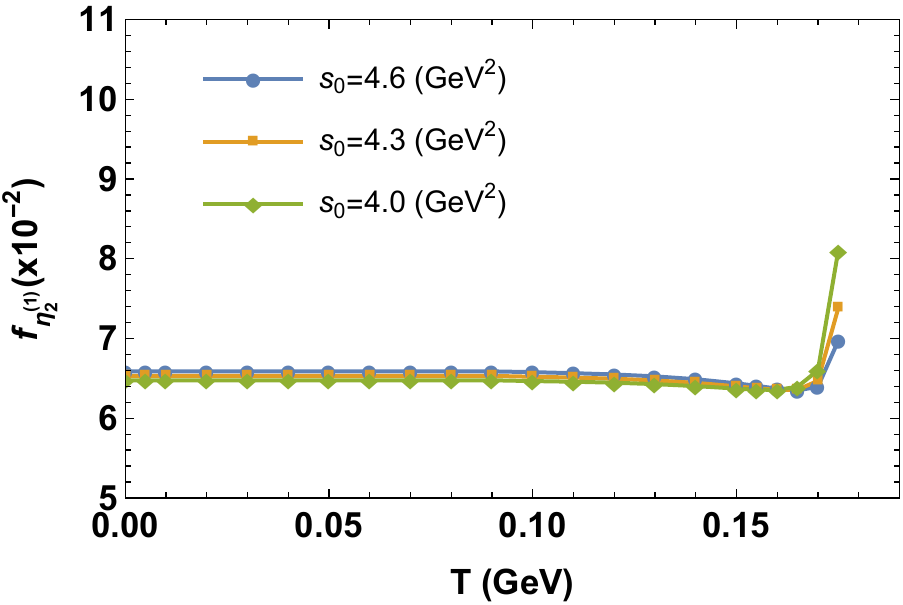}\hskip0.1cm
\includegraphics[totalheight=5.2cm,width=5.8cm]{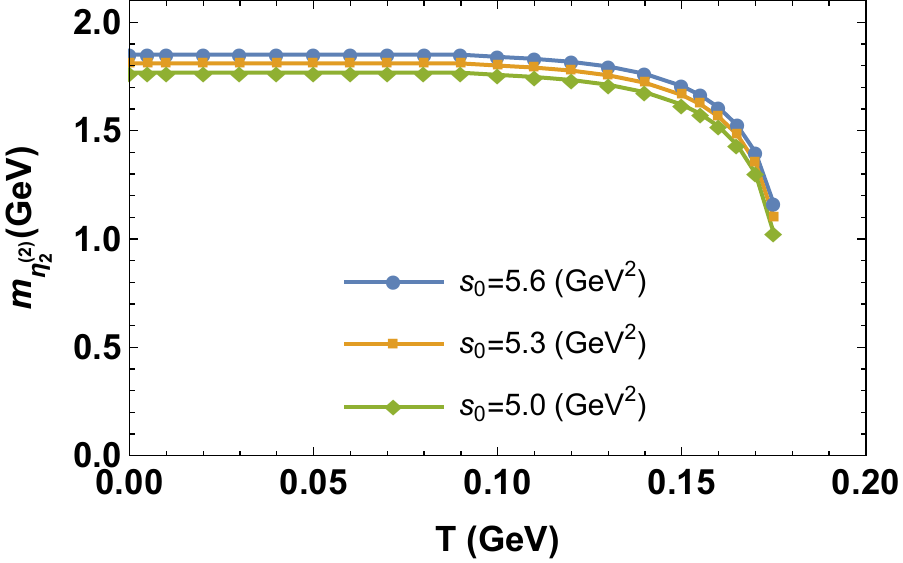}\hskip0.1cm
\includegraphics[totalheight=5.2cm,width=5.8cm]{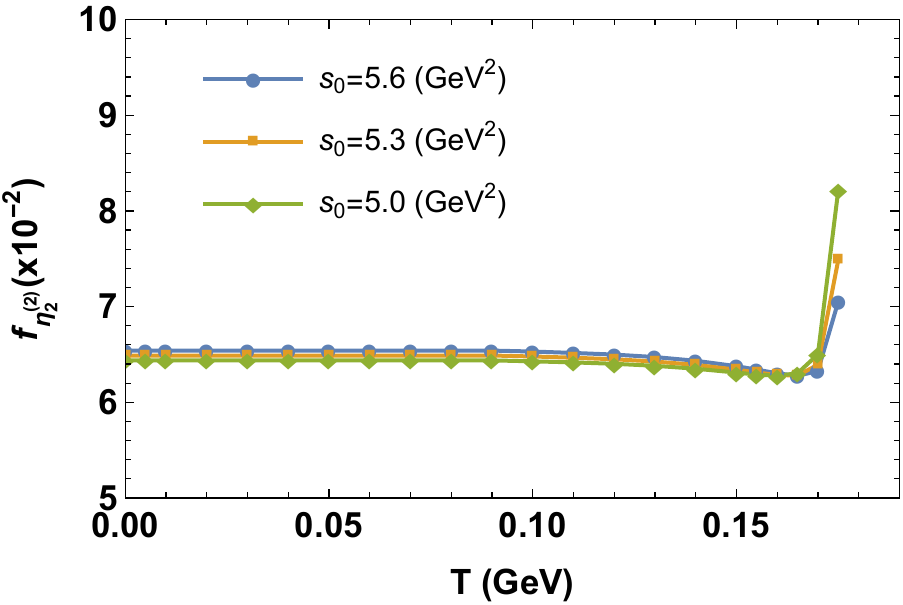}
\end{center} 
\caption{The masses and decay constants of the $\pi_2(1670),\eta_2(1645) $ and $\eta_2(1870)$ as a function of temperature at some fixed values of continuum threshold parameter $s_0$.}\label{figmassandf}
\end{figure}
Our numerical results depict that the mass values of the $\pi_2(1670)$, $\eta_2(1645)$ and $\eta_2(1870)$ states are stable at low temperatures, but they decrease approximately by $49\%$, $38\%$ and $39\%$ at above nearly critical temperature as shown in Figure~\ref{figmassandf}. Meanwhile decay constant value of $\pi_2(1670)$ drops by $10\%$, but $f_{\eta_2(1645)}$ and $f_{\eta_2(1870)}$ increases by $13\%$, $15\%$, respectively when analysed up to  $T=180~ \mathrm {MeV} $.  These changes may be signal of deconfinement phase transition in hot medium.
%
%%%%%%%%%%%%%%%%%%%%%%%%%%%%%%%%%%%%%%%%%%%%%%%%%%%%%%%%%%%%%%%%%%%%%%%%%%%%%
\section{Summary}\label{sec:Result}
%%%%%%%%%%%%%%%%%%%%%%%%%%%%%%%%%%%%%%%%%%%%%%%%%%%%%%%%%%%%%%%%%%%%%%%%%%%%%
	
	Analysing hadrons at finite temperatures is the popular current research area to explore the QGP which is thought to be available in the early stages of the universe and possibly in the inner core of neutron stars. In this context, surveying the hadrons in hot medium can provide precious hints on the dynamics of QCD.  In a hot bath, we can probe the changes of QCD vacuum, by examining the thermal effects on the condensates and observing the melting of hadrons at estimated deconfined temperatures. Investigations lead to the results consistent with the deconfinement temperature  $T_c=155~\mathrm{MeV}$~\cite{Aoki:2006br}-\cite{Bazavov:2017dus}. 
	
	In this paper, we explored the pseudotensor mesons $\pi_2(1670)$, $\eta_2(1645)$ and $\eta_2(1870)$ in hot bath. To describe the effects of hot medium on the hadronic parameters of these resonances, the Thermal QCD sum rule model is employed taking into account the contributions of condensates up to dimension five. In calculations, we assume that the OPE and quark-hadron duality stay valid at finite temperatures, however, we substitute the vacuum condensates by their thermal expectation values. In addition, our data give approximately the same values as our previous work, which we made using a different fit function \cite{Sungu:2019imas,Turkan:2019imas} and pseudotensor current, our results at $T=0$ limit are also in reasonable agreement with the available experimental data.
	
	A remarkable changes in the values of mass and decay constant in hot medium can be interpreted as the signal of QGP, called a new state
	of matter. Also, it might indicate the deconfinement phase transition in QGP which
	may occur in the primordial universe. So the behaviour of pseudotensor mesons
	in terms of temperature can be a useful tool to analyse the
	heavy-ion collision experiments. We hope that our predictions for the hadronic features of the considered mesons in hot medium can be tested in near future
	experiments.
	\appendix
	\section{Mesonic Spectral Functions at Finite Temperature}\label{sec:App}
	The explicit form of the perturbative part of the spectral
	density in Eq.~(\ref{eq:MassSR}) and Eq.~(\ref{eq:DecConSR}) is found as for the $\pi_2$ meson:
	\begin{eqnarray}\label{eq:rhopertpi2}
		\rho^{{Pert.}}(s)=\frac{[10 (m_d^2 +m_u^2) - 9 s]s} 
		{240 \pi^2 }.
	\end{eqnarray}
	The nonperturbative contributions in Eq.~(\ref{eq:MassSR}) and Eq.~(\ref{eq:DecConSR}) is evaluated via the Borel transformation and has the following form, respectively:
	\begin{eqnarray}\label{Dim3Pi2}
B\Pi^{\langle q\bar{q}\rangle}(q^2,T)&=&-\frac{1}{3} M^2 \Big(\langle \bar{d}d \rangle m_d + m_u \langle \bar{u}u \rangle \Big),
	\end{eqnarray}
\begin{eqnarray}\label{Dim4Pi2}
B\Pi^{\mathrm{\langle G^2 \rangle+\langle \Theta_{00}\rangle}}(q^2,T)&=&0,
\end{eqnarray}
\begin{eqnarray}\label{Dim5Pi2}
B\Pi^{\langle \bar{q}G q\rangle}(q^2,T)&=& \frac{1}{4} m_0^2 \Big(\langle \bar{d}d \rangle  m_d + m_u \langle \bar{u}u \rangle \Big),
\end{eqnarray}
and for the $\eta_2(1645)$ as follows:
\begin{eqnarray}\label{eq:rhoperteta1645}
\rho^{{Pert.}}(s)&=&\frac{[20 (m_d^2 + m_s^2 + m_u^2) - 27s] s} {720 \pi^2},
\end{eqnarray}
\begin{eqnarray}\label{Dim3eta1645}
B\Pi^{\langle q\bar{q}\rangle}(q^2,T)&=&-\frac{2}{9} M^2 \Big(\langle \bar{d}d \rangle m_d + m_s \langle \bar{s}s \rangle +m_u\langle \bar{u}u \rangle \Big),
\end{eqnarray}
\begin{eqnarray}\label{Dim4eta1645}
B\Pi^{\mathrm{\langle G^2 \rangle+\langle \Theta_{00}\rangle}}(q^2,T)&=&0,
\end{eqnarray}
%%%%%%%%%%%%%%%%%%%%%%%%%%%%%%%%%%%%%%%%%%%%%%%%%%%%%%%%%%%%%%%%%%%%%%%%%%%%%%
\begin{eqnarray}\label{Dim5eta1645}
B\Pi^{\langle \bar{q}G q\rangle}(q^2,T)&=& \frac{5}{18} m_0^2 \Big(\langle \bar{d}d \rangle  m_d + m_s \langle \bar{s}s \rangle +m_u \langle \bar{u}u \rangle\Big),
\end{eqnarray}
and also for the $\eta_2(1870)$ we get the below expressions:
%%%%%%%%%%%%%%%%%%%%%%%%%%%%%%%%%%%%%%%%%%%%%%%%%%%%%%%%%%%%%%%%%%%%%%%%%%%%
\begin{eqnarray}\label{eq:rhoperteta1870}
	\rho^{{Pert.}}(s)&=&\frac{(10 m_d^2 + 40 m_s^2 + 10 m_u^2 - 27 s) s}{
		720 \pi^2},
\end{eqnarray}
\begin{eqnarray}\label{Dim3eta11870}
	B\Pi^{\langle q\bar{q}\rangle}(q^2,T)&=&-\frac{1}{9} M^2 \Big(\langle \bar{d}d \rangle m_d + 4m_s \langle \bar{s}s \rangle +m_u\langle \bar{u}u \rangle \Big),
\end{eqnarray}
\begin{eqnarray}\label{Dim4eta1870}
	B\Pi^{\mathrm{\langle G^2 \rangle+\langle \Theta_{00}\rangle}}(q^2,T)&=&0,
\end{eqnarray}
\begin{eqnarray}\label{Dim5eta1870}
	B\Pi^{\langle \bar{q}G q\rangle}(q^2,T)&=& \frac{5}{36} m_0^2 \Big(\langle \bar{d}d \rangle  m_d +4 m_s \langle \bar{s}s \rangle +m_u \langle \bar{u}u \rangle\Big)
\end{eqnarray}
%%%%%%%%%%%%%%%%%%%%%%%%%%%%%%%%%%%%%%%%%%%%%%%%%%%%%%%%%%%%%%%%%%%%%%%%%%%%%
with $  m^2_0=(0.8\pm0.2)~\mathrm{GeV}^2 $

\end{document}